\documentclass[10pt,letterpaper,twocolumn]{article} 

\usepackage{ol2}
\usepackage[draft,implicit=false]{hyperref}
\usepackage{amsmath}

\begin{document}

\twocolumn[ 

\title{Broadband active tuning of unidirectional scattering from nanoantenna using combined radially and azimuthally polarized beams}


\author{Zheng Xi,$^{1,*}$ Lei Wei,$^1$ A.J.L. Adam,$^1$, H.P. Urbach,${^1,^2}$}

\address{
$^1$ Optics Research Group, Department of Imaging Physics, Faculty of Applied Sciences \\ Delft
University of Technology, Van der Waalsweg 8, 2628CH Delft, The Netherlands \\
$^2$  H.P.Urbach@tudelft.nl\\

$^*$Corresponding author: Z.Xi@tudelft.nl
}

\begin{abstract}We propose an approach to actively tune the scattering pattern of a Mie-type spherical antenna. The scheme is based on separate control over the induced electric dipole and induced magnetic dipole using two coherent focused beams of radial polarization and azimuthal polarization. By carefully tuning the amplitude and phase relation of the two beams, a broadband unidirectional scattering can be achieved, even at the wavelength where the antenna scatters most efficiently. By moving the focus of one beam, a drastic switch of the unidirectional scattering can be observed. Such scheme enables the design of ultra-compact optical switches and directional couplers based on nanoantennas.  \end{abstract}


 ] 

\noindent 
The study of the scattering of light by nanoparticles has been a topic of great interest in the field of wireless communication at nanoscale, imaging of biological samples and detection of subwavelength defects\cite{Bharadwaj2009,Novotny2011}. Among them, much attention has been paid to achieve unidirectional scattering of light because of its importance in the designing of optical antennas to concentrate and direct light at nanoscale with applications ranging from cloaked SNOM probe to collect signals\cite{Alu2009}, nanoscale optical switch for waveguide\cite{2040-8986-16-10-105002}, photovoltaic absorbers with high efficiency\cite{atwater2010plasmonics}.

Recently, it has been shown that by balancing the electric dipole response and magnetic dipole response inside high refractive dielectric nanoantenna structures, a unidirectional scattering behaviour can be observed\cite{Person2013a,Fu2013,geffrin2012magnetic}. This unidirectional scattering behaviour is strongly wavelength dependent. For a certain wavelength, where the electric and magnetic response are equal and in phase, the nanoantenna can act as a “Huygens source,” with most of the energy been scattered in the forward direction. Whereas for another wavelength, the electric and magnetic responses are equal but nearly out of phase, the nanoantenna can scatter most of the energy in the backward direction. However, because the electric and magnetic responses of the antenna are highly wavelength dependent, the unidirectional scattering schemes shown previously are narrowband. Besides, the wavelength at which the two responses are equal is not the wavelength for which the antenna scatters most efficiently, it’s often an off-resonant scattering. Moreover, the scattering direction cannot be actively changed without modifying the material property of the sphere, which hinders its application in nanoscale active switch. It is the intention of this Letter to show that by independently exciting the electric and magnetic dipolar resonances inside the antenna using two focused coherent beams of radial and azimuthal polarization, the unidirectional scattering can be achieved within a broad wavelength range, including the wavelength corresponding to the strongest scattering of the nanoantenna. Besides, we show that the scattering direction can be actively flipped by changing position of the focal point of one beam. 

\begin{figure}[htbp]
\centering
{\includegraphics[width=\linewidth]{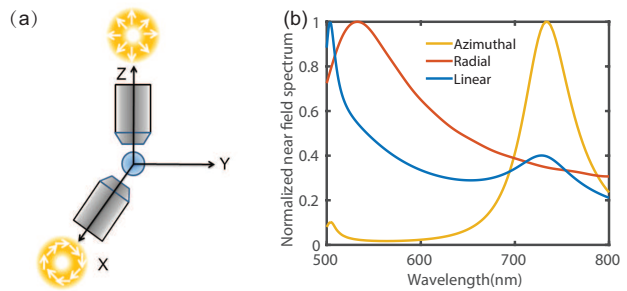}}
\caption{(a) Configuration of the scheme, a spherical antenna of radius 100~nm with refractive index 3.5 is placed at the centre of the coordinate system. An azimuthally polarized beam and a radially polarized beam are focused onto the spherical antenna with NA=0.86 (b) Near field spectrum obtained by summing up electric field intensity inside the nanoantenna using linearly (blue curve), radially (red curve) and azimuthally (yellow curve) polarized light.}
\label{fig:false-color}
\end{figure}

The configuration of our proposed scheme is shown in Figure 1(a). With no loss of generality, the nano sphere antenna considered here is of refractive index n=3.5 and radius r=100 nm, but the scheme can be applied to other high refractive index Mie-type particles as well. The surrounding medium is assumed to be air.  In contrast to previous studies which involve only one linear polarized beam as the excitation beam, here two perpendicularly oriented focused laser beams along X and Z axis are used. The polarizations of the two beams are chosen to be radial and azimuthal polarization respectively. The focused beam profiles are calculated at a defocused plane using Richard-Wolf diffraction integral which can be imported into finite difference time domain simulation as the excitation beams\cite{Novotny2012}. In order to achieve accurate field distribution at the focus, a beam area of 10 $\mu$m by 10 $\mu$m is used. When focused, the radial and azimuthal polarization exhibit pure longitudinal electric $E_{z}$ along Z axis and magnetic $H_{x}$ field along X axis at the focus respectively. If a nano sphere antenna is placed at the focus of the focal beam, unlike the case for linear polarized light where both the incident electric and magnetic field are coupled to the antenna, it can only interact with either the electric or the magnetic field component of the focal field. As a result, when the nano sphere antenna is illuminated with focused radially or azimuthally polarized beams, some of its resonances can be switched on and off as shown in Fig. 1(b)\cite{wozniak2015selective}. The near field enhancement spectrum obtained by summing up field intensity inside the antenna reveals that the electric dipolar resonance at 533 nm can only be excited by focused radially polarized light, whereas the magnetic dipolar and quadruple resonances at 734 nm and 505 nm respectively can only be excited by focused azimuthally polarized light. This excitation scheme provides us more flexibility in selective excitation of the electric and magnetic resonances of the antenna.

\begin{figure}[htbp]
\centering
{\includegraphics[width=\linewidth]{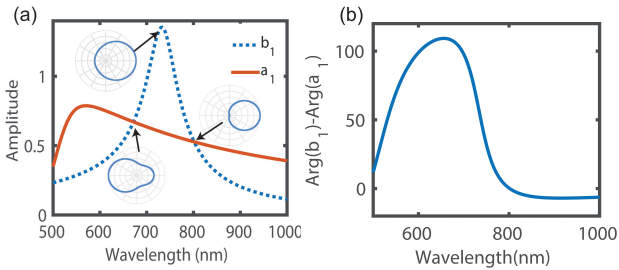}}
\caption{(a)The amplitude and (b) the phase difference (in degree) of the first two Mie coefficients ${a}_{1}$ and ${b}_{1}$. Insets show the farfield scattering power pattern under linear polarized light excitation along +Y axis.}
\label{fig:false-color}
\end{figure}

When the sphere is small enough compared to the wavelength, its far field scattering response to the incident field can be fully described using the radiation of the two dominant induced dipole resonances ${{p}_{induced}}={{\epsilon }_{0}}{{\alpha }_{e}}{{E}_{in}}$ and ${{m}_{induced}}={{\alpha }_{m}}{{H}_{in}}$, where ${E}_{in}$ and ${H}_{in}$ are the fields that polarize the sphere. According to Mie theory, the electric dipolar polarizability and magnetic dipolar polarizability can be written as ${{\alpha }_{e}}=i\frac{6\pi }{{{k}^{3}}}{{a}_{1}}$ and ${{\alpha }_{m}}=i\frac{6\pi }{{{k}^{3}}}{{b}_{1}}$ with ${{a}_{1}}$ and ${{b}_{1}}$ representing the first two Mie coefficients. For the spherical antenna discussed here, we plot in Figure 2 the first two dipolar polarizabilities ${{a}_{1}}$ and ${{b}_{1}}$, for both amplitude and phase difference. The red solid curve in Figure 2(a) shows the contribution of the electric dipolar term ${a}_{1}$ to the scattered far field while the blue dashed curve is the contribution from the magnetic dipole ${b}_{1}$. The two peaks at 560 nm and 732 nm are the resonances of the antenna due to the electric and magnetic dipolar terms respectively. The resonances at the far-field are shifted with respect to the maximum of near field enhancement\cite{zuloaga2011energy}. At 802 nm, the amplitude of the two dipole responses are equal and the phase difference between them is 0. At this point, the induced electric dipole and magnetic dipole moments satisfy the condition $cp_{induced}=m_{induced}$ which is known as the first Kerker’s condition\cite{Alaee2015,Kerker1983}, the interference of the two induced dipoles leads to zero backward scattering as shown in the inset. At 672 nm, the second Kerker’s condition is satisfied\cite{Alaee2015,Kerker1983}, although the amplitudes of the two are the same, the phase difference between them is $108^o$, the destructive interference of the two dipoles is not complete in the forward direction. Therefore a small tail can be seen in the forward direction. It is also worth mentioning that at 732 nm, although the particle scatters most efficiently to the incident light, the scattering pattern shows only slight asymmetry. However, an inspection to the expression of the induced dipoles reveals that they are not only dependent on the polarizability, but can also be affected by the electric and magnetic field component ${E}_{in}$ and ${H}_{in}$. As mentioned above, the radial and azimuthal beams can excite the electric and magnetic dipole terms respectively without affecting the other. If the amplitude and the phase of the two beams at the center of the focus are chosen such that they fulfill the following conditions

\begin{align}
   \nonumber
	 |{{E}^{rad}_{focus}}|/|{{H}^{azi}_{focus}}|&=Z_{0}|{{b}_{1}}|/|{{a}_{1}}| \\  
   \arg ({{E}^{rad}_{focus}})-\arg({{H}^{azi}_{focus}})&=\arg ({{b}_{1}})-\arg({{a}_{1}})\ \\ \nonumber
     or \arg ({{E}^{rad}_{focus}})-\arg({{H}^{azi}_{focus}})&= 180 -(\arg ({{b}_{1}})-\arg({{a}_{1}})\ ) \\  \nonumber
\end{align}
the mismatch due to the polarizability of the sphere can be compensated by the incident light, a unidirectional scattering can always be expected. From an experimental point of view, it is more straightfoward to control the incoming field before the focusing objectives, therefore we show next the relationship between the input radial and azimuthal fields. According to Richard-Wolf diffraction integral, the field at the focal point of radially polarized beam only contains the longitudinal electric field component, in the configuration discussed above, ${E}^{rad}_{focus}=AE^{rad}_{in}I_{10}$ and for an azimuthally polarized beam, it only contains the longitudinal magnetic field ${H}^{azi}_{focus}=\frac{AE^{azi}_{in}I_{10}}{Z_{0}}$ where $A$ is a constant for a certain focusing system and $I_{10}$ is a fixed value at the focal point, $E^{azi}_{in}$ and $E^{azi}_{in}$ are the complex amplitudes of the two incoming beams\cite{Novotny2012}. Note here, the field at the center of the focus ${E}^{rad}_{focus}$ and ${H}^{azi}_{focus}$ are directly proportional to the incoming field $E^{rad}_{in}$ and $E^{azi}_{in}$, the above unidirectional scattering condition for the two incoming beams can be rewritten as 

\begin{align}
   \nonumber
	 |{{E}^{rad}_{in}}|/|{{E}^{azi}_{in}}|&=|{{b}_{1}}|/|{{a}_{1}}| \\  
   \arg ({{E}^{rad}_{in}})-\arg({{E}^{azi}_{in}})&=\arg ({{b}_{1}})-\arg({{a}_{1}})\ \\ \nonumber
     or \arg ({{E}^{rad}_{in}})-\arg({{E}^{azi}_{in}})&= 180 -(\arg ({{b}_{1}})-\arg({{a}_{1}})\ ) \\  \nonumber
\end{align}
As we will see in the following, this tuning of the property of the beams themselves serves as a robust way of tuning the scattering property of the nano spherical antenna.

\begin{figure}
\centering
{\includegraphics[width=\linewidth]{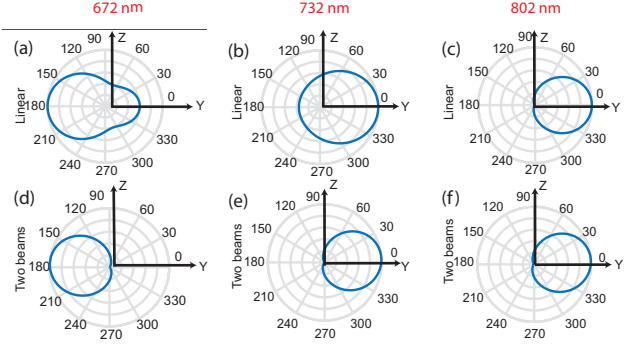}}
\caption{The far-field scattering power pattern (a),(d) 672 nm, (b),(e) 732 nm and (c),(f) 802 nm. The upper row corresponds to linear polarization excitation along +Y direction and the lower row corresponds to pattern obtained by two beams excitation.}
\label{fig:false-color}
\end{figure}

In Figure 3, we plot the far-field scattering power pattern in the YZ plane at the wavelength of 672 nm, 732 nm and 802~nm using two beams as excitation with relative amplitude and phase fulfilling Eq. (2). For comparison, the far-field scattering power pattern at these three wavelengths for linear polarization excitation along +Y direction are plotted. The little asymmetry in the far-field plane with respect to the Y axis for the two beams case is due to asymmetric phase distribution of the incident beam as it propagates through the focus. At 802 nm, the amplitude of the two beams are equal and the phase difference is 0, because the polarizabilities already provide the required amplitude and phase difference, the far-field scattering lies mainly in +Y direction just like in the linear polarization case. At 672 nm, the amplitude of the two polarizabilities are the same but the phase difference is $108^{o}$. Therefore the amplitude ratio of the two beams are chosen to be $|{{E}^{rad}_{in}}|/|{{E}^{azi}_{in}}|=1$ and $\arg ({{E}^{rad}_{in}})-\arg({{E}^{azi}_{in}})={{72}^{o}}$ to compensate the phase difference in the induced dipoles by polarizabilities. A complete suppression of the small tail in the unwanted direction can be achieved. The scattering at 732 nm is of special interest, because at this point, the antenna scatters most efficiently. However, it can be seen from Figure 2(a) that the dominant contribution to the scattered field is from $b_{1}$ which is the magnetic dipolar term and this term shows a uniform distribution at the considered far field plane. By adding a small contribution from the electric dipole term $a_1$, the far field scattering pattern becomes slightly asymmetrical. However, if the ratio between two excitation beams is set to be $|{{E}^{rad}_{in}}|/|{{E}^{azi}_{in}}|=2.27$ and $\arg ({{E}^{rad}_{in}})-\arg({{E}^{azi}_{in}})=-{{59.4}^{o}}$ according to Eq. (2) to enhance the strength of the induced electric dipole and bridge the mismatch of the phase. A unidirectional scattering pattern can be observed even at this mostly scattered wavelength. Besides, if the phase difference between the two beams is added by $180^{o}$, then the scattering direction can be switched accordingly. This can be used as an efficient optical switch at nanoscale.

\begin{figure}[htbp]
\centering
{\includegraphics[width=\linewidth]{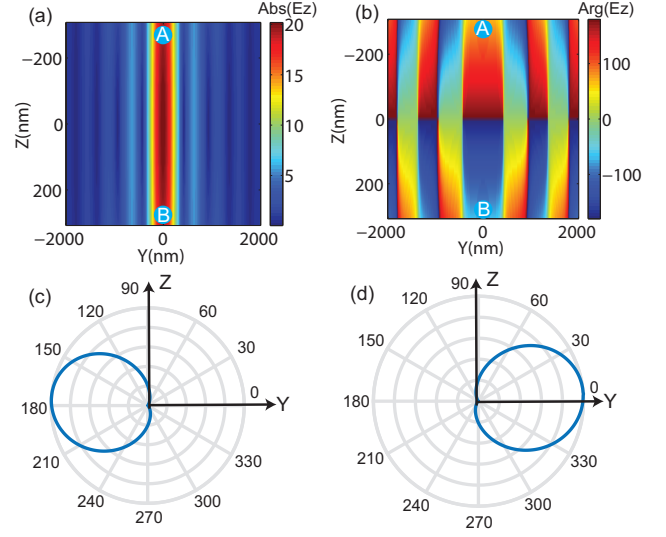}}
\caption{(a) and (b) Amplitude and phase distribution (in degree) along the focal axis of a focused radially polarized beam. (c) and (d) Far field scattering power pattern at these two points.}
\label{fig:false-color}
\end{figure}

The above tuning scheme can be readily implemented using existing spatial light modulators to control the amplitude and the phase of the two beams originated from one single source. Additionally, a different tuning scheme can be proposed if the field distribution along one beam’s focal axis is considered. In Figure 4 (a) and (b), we plot the amplitude and the phase distribution of the longitudinal field component $E_{z}$ along the focal axis of the radially polarized beam. The wavelength of the incident light is 732nm and the numerical aperture is 0.86. At point A and point B, the $E_{z}$ fields are of the same amplitude but oscillate out of phase. According to Eq. (2), if the condition for unidirectional scattering at point A is satisfied, then if the focal position is changed such that the antenna originally located at point A is moved to the point B, because the $E_{z}$ fields at these two points only vary by the phase difference of $180^{o}$ with the same amplitude, the direction of scattering can be switched by $180^o$ as shown in Figure 4(c) and (d). The amplitude and phase ratio chosen here is $|{{E}^{rad}_{in}}|/|{{E}^{azi}_{in}}|=2.68$ and $\arg ({{E}^{rad}_{in}})-\arg({{E}^{azi}_{in}})={{31.63}^{o}}$ to meet unidirectional scattering condition at point A. This scheme can be further used to probe the near field phase information of the focal field which can only be accessed by complicated interferomety method\cite{bauer2014nanointerferometric}. 

Finally, it is important to emphasize again that the two beams scheme does not involve the tuning of the material or the structure of the sphere to achieve broadband unidirectional scattering. By carefully engineering the relative amplitude and phase relation between two focused radially and azimuthally polarized beams, a broadband unidirectional scattering property can be achieved as long as the first two dipolar terms of the Mie theory dominate. The proposed scheme can also be applied to more complicated structures involving higher multiple resonances as well to achieve higher directivity\cite{Alaee2015}.

In summary,  we have proposed a versatile scheme to achieve controllable unidirectional scattering of a nano sphere antenna within a broad bandwidth. By carefully engineer the amplitude and phase of the two incident beams, the individual induced dipoles can be tuned independently so that the condition for unidirectional scattering can always be satisfied within a broad range. By changing the position of the antenna with respect to the focal point of one beam, a directional switch of the far field scattering can be realized.










\newpage
\null
\newpage
\null

\end{document}